\documentclass[preprint,aps,showpacs,nofootinbib]{revtex4}

\def\bea{\begin{eqnarray}}
\def\eea{\end{eqnarray}}
\def\bec{\begin{center}}
\def\ec{\end{center}}
\def\beq{\begin{equation}}
\def\eeq{\end{equation}}

\begin{document}

\draft
\tighten
\preprint{KIAS-P08066}
\title{\large \bf Moduli stabilization with F-term uplifting
\\
in heterotic string theory}
\author{
Kwang Sik Jeong\footnote{email: ksjeong@kias.re.kr}$^1$
and Seodong Shin\footnote{email: sshin@muon.kaist.ac.kr}$^2$}
\affiliation{
$^1$School of Physics, Korea Institute for Advanced Study,
Seoul 130-722, Korea \\
$^2$Department of Physics, KAIST, Daejeon 305-701, Korea}

\vspace{1cm}

\begin{abstract}
We discuss the role of F-term uplifting in stabilizing moduli within the
framework of heterotic string theory.
It turns out that the uplifting sector plays an important role in fixing the
volume modulus at one of the self-dual points of a modular invariant potential.
For the volume modulus stabilized at a self-dual point, the F-term uplifting
leads to the dilation stabilization which can naturally yield the mirage
mediation pattern of soft supersymmetry breaking terms.
Generalizing to the case with anomalous $U(1)$ gauge symmetry, we also find
that the $U(1)$ sector generically gives a contribution to sfermion masses
comparable to the dilaton mediated one while maintaining the mirage mediation
pattern.
\end{abstract}

\maketitle

\section{Introduction}

In supergravity theories derived from string compactification, effective
couplings of gauge and matter superfields are determined by the vacuum values
of string moduli including the dilaton.
To fix moduli vacuum values, one can incorporate various
non-perturbative effects and background fluxes which generate a non-trivial
potential for the moduli.
However, it is usually argued that such sources of the moduli potential,
though responsible for stabilizing the moduli, are alone incapable of
rendering the vacuum energy adjustable to a small positive value.
Typically there arises a moduli vacuum solution with negative vacuum energy.
This has led to consider the construction of a realistic de-Sitter vacuum via
an uplifting mechanism \cite{Kachru:2003aw} that is implemented by adding
extra supersymmetry breaking effects.
Since the uplifting effects generally shift the moduli vacuum configuration,
how the moduli break supersymmetry would depend crucially on the uplifting
procedure \cite{Choi:2004sx}.
The uplifting mechanism has been studied mostly within the framework of
type IIB string theory where the use of fluxes is rather flexible
\cite{Kachru:2003aw,Choi:2004sx,GomezReino:2006dk}.
Such moduli stabilization has been noticed to provide a natural scheme realizing
the mirage mediation of supersymmetry breaking \cite{Choi:2005uz},
in which anomaly mediation is equally important as moduli mediation.

Recently L\"owen and Nilles have discussed the possibility of F-term uplifting
scenario within the framework of heterotic string theory \cite{Lowen:2008fm}.
They considered a moduli potential which is non-perturbatively generated by
gaugino condensation in a modular invariant effective supergravity.
Then, introducing F-term uplifting sector, they found that the heterotic dilaton
can be successfully stabilized with a single gaugino condensation in a
controllable approximation while achieving the desired value of the vacuum
energy.
This is because the uplifting sector plays an important role in fixing the
dilaton as well as in adjusting the vacuum energy.
In their discussion, the volume modulus in heterotic compactifications is
assumed to be stabilized at one of the self-dual points of a modular invariant
potential, giving no contributions to the supersymmetry breakdown.

Since the uplifting matter field generically has non-trivial modular
transformation, uplifting effects might shift the volume modulus from
the self-dual point.
Furthermore, it has been noticed that the self-dual points typically correspond
to local maxima or saddle points in the absence of uplifting sector.
Motivated by these issues, we wish to examine in more detail the role of F-term
uplifting sector and the stabilization of moduli in heterotic compactifications
while explicitly including the volume modulus.
It turns out that the inclusion of uplifting sector is indeed crucial for
stabilizing the volume modulus at one of the self-dual points.
To see this, we derive the condition for a self-dual point to be a local minimum
of modular invariant potential, from which the modular weight of uplifting field
is found to be somewhat constrained.

Fixed at a self-dual point, the volume modulus does not participate in
supersymmetry breaking.
The F-term uplifting scheme then leads to the dilation stabilization that
can naturally realize the mirage mediation scenario where anomaly and dilaton
mediations give comparable contributions to soft supersymmetry breaking terms.
Meanwhile, heterotic compactifications often involve anomalous $U(1)$ gauge
symmetries, under which the dilaton transforms non-linearly to implement the
Green-Schwarz mechanism of anomaly cancellation \cite{Green:1984sg}.
Generalizing the F-term uplifting scheme to a situation involving anomalous
$U(1)$ gauge symmetry, we will also examine its implications for supersymmetry
breaking.
It is found that the $U(1)$ sector gives a comparable contribution to sfermion
soft terms as the dilaton mediation \cite{Choi:2006bh}
while maintaining the mirage pattern.

This paper is organized as following.
In section 2, we shall discuss how the potential for heterotic moduli is
affected by F-term uplifting sector, and then examine the condition for the
volume modulus to be stabilized at one of the self-dual points.
Section 3 contains a brief analysis of the pattern of soft terms.
Extending to a more general case with anomalous $U(1)$ gauge symmetry,
the implications of $U(1)$ sector on soft terms will be examined in section 4.
Section 5 is the conclusion.

\section{Moduli potential and F-term uplifting}
To describe the low energy dynamics of heterotic string theory, we consider
the effective four-dimensional supergravity under the assumption that
the moduli dependence of K\"ahler potential is well approximated by the string
tree-level expression:
\bea
\label{K}
K &=& -3\ln(T+\bar T) - \ln(S+\bar S),
\eea
where $T$ is the volume modulus and $S$ is the dilaton.
The supergravity action is specified by the K\"ahler invariant function
$G=K+\ln|W|^2$, where $W$ is the holomorphic superpotential.
For the above tree-level K\"ahler potential, the effective superpotential is
required to have explicit dependence on both $T$ and $S$ in order to generate
a non-trivial potential for their stabilization.
Since the tree-level gauge kinetic function is given by $f_a=S$ for all gauge
groups ${\cal G}_a$, gaugino condensation \cite{Ferrara:1982qs}
can naturally generate non-perturbative superpotential for the dilaton.
As for the volume modulus, a dependence on $T$ is introduced
in the superpotential if one requires the modular invariance of effective
supergravity action that arises in orbifold compactifications
\cite{Ferrara:1989bc}.
Considering such sources of moduli dependence, it follows that
the superpotential generically has a moduli dependence of the form
\bea
\label{W}
W &=& \frac{1}{\eta(T)^6}\omega(S),
\eea
which can be inferred from the modular invariance of $G$.
Here $\eta(T)=e^{-\pi T/12}\prod_n(1-e^{-2\pi n T})$ is the Dedekind eta
function.
With this superpotential involving $T$ and $S$, it is now possible to provide
a potential for the moduli.

Since there is no mixing between the dilaton and the volume modulus in the
tree-level K\"ahler potential, the scalar potential reads
\bea
V &=& e^G\left(G^{-1}_{I\bar J}G_IG_{\bar J} -3 \right)
= e^G\left(\frac{1}{G_{T\bar T}}|G_T|^2 + \frac{1}{G_{S\bar S}}|G_S|^2
- 3 \right),
\eea
where $G_I\equiv\partial_I G$, and $G^{-1}_{I\bar J}=K^{-1}_{I\bar J}$
denotes the inverse K\"ahler metric.
This potential should have a minimum at an acceptable value for the dilaton
since the gauge coupling constant is given by $1/g^2={\rm Re}(S)$.
The vacuum value of $S$ is determined by solving the extremum condition
$\partial_S V=0$:
\bea
\frac{G_{SS\bar S}}{G^2_{S\bar S}}|G_S|^2
- (1+e^{-G}V)G_S - \frac{G_{SS}}{G_{S\bar S}}G_{\bar S}
&=& \frac{G_{TS}}{G_{T\bar T}}G_{\bar T},
\eea
in which $G_{TS}=0$ for the modular invariant action derived from (\ref{K})
and (\ref{W}).
Hence there is a possible solution at $G_S=0$ regardless of the value of
$G_T$ which is proportional to the F-term of $T$ \cite{de Carlos:1992da}.
Multiple gaugino condensations\footnote{For a single gaugino condensation
$\omega(S)=A e^{-a S}$, the potential runs away as $S$ goes to infinity.}
indeed lead to the potential which can develop a minimum along the $S$-direction
at $G_S=0$ through the racetrack mechanism \cite{Krasnikov:1987jj}.
On the other hand, it is quite difficult to achieve a solution with
non-vanishing $G_S$ for ${\rm Re}(S)={\cal O}(1)$ because the typical size
of $G_{SS}$ is $|G_{SS}|\gg 1$ if the dilaton superpotential arises
non-perturbatively from gaugino condensation.

Meanwhile, the modular invariance of supergravity action ensures that
the self-dual points $T=1,e^{i\pi/6}$ always correspond to extrema of
the potential where the F-term of volume modulus is vanishing.
To stabilize the volume modulus at one of the self-dual points,
the condition for a local minimum requires
\bea
\partial_T\partial_{\bar T} V &>& |\partial^2_T V|,
\eea
where we have used $\partial_T\partial_S V=\partial_T\partial_{\bar S} V=0$
at the self-dual points.
However, for the moduli potential derived from (\ref{K}) and (\ref{W}),
this requirement cannot be fulfilled as long as the dilaton is stabilized
at $G_S=0$.
Actually the potential develops a minimum in $T$ close to the point $T=1.2$
\cite{Font:1990nt},
which corresponds to a deep anti de-Sitter with non-vanishing F-term for $T$.
Although the moduli $T$ and $S$ can both be stabilized  reliably, there still
remains a problem since it is difficult to construct a realistic de-Sitter
vacuum consistent with the cosmological observations.
Therefore one has to assume the existence of some contributions compensating
the negative vacuum energy.

To get a desired value of the vacuum energy, L\"owen and Nilles have considered
F-term uplifting mechanism under the assumption that the volume modulus is
stabilized at one of the self-dual points with vanishing F-term
\cite{Lowen:2008fm}.
They found that the uplifting sector plays an important role not only in
adjusting the vacuum energy but also in achieving the stabilization of dilaton.
In this work, we examine in detail the role of uplifting sector while
explicitly including the volume modulus.
Including the uplifting matter field $Z$, the effective K\"ahler potential
is now written as
\bea
\label{KF}
K &=& -3\ln(T+\bar T) - \ln(S+\bar S) + \frac{|Z|^2}{(T+\bar T)^{n_Z}},
\eea
where $n_Z$ denotes the modular weight of $Z$, which is generically a rational
number of order unity.
For the supergravity action invariant under modular transformations,
the modular invariance of $G$ leads to the relations
\bea
\partial_T V &\propto& n_Z Z \partial_Z V,
\nonumber \\
F^T &\propto& G_{Z\bar Z}G_{\bar T} - G_{\bar T Z}G_{\bar Z} = 0,
\eea
at the self-dual points $T=1,e^{i\pi/6}$.
It is therefore obvious that extrema of the potential still occur along the
$T$-direction at the self-dual points where its F-term does vanish.
Since the uplifting sector gives an additional contribution to the potential for
the volume modulus,
it is necessary to reexamine the possibility that the self-dual points can
correspond to a local minimum of the potential.

If the volume modulus is stabilized at one of self-dual points, it has a
vanishing F-term and thus the scalar potential is written as
\bea
V &=& e^{G}\left(\frac{1}{G_{S\bar S}}|G_S|^2
+ \frac{1}{G_{Z\bar Z}}|G_Z|^2 - 3 \right),
\eea
and the stationary condition for the dilaton is given by
\bea
\label{VS0}
\frac{G_{SS\bar S}}{G^2_{S\bar S}}|G_S|^2
- (1+e^{-G}V)G_S - \frac{G_{SS}}{G_{S\bar S}}G_{\bar S}
&=&
\frac{G_{S Z}}{G_{Z\bar Z}}G_{\bar Z},
\eea
from which the dilaton should be fixed in a realistic region with
${\rm Re}(S)={\cal O}(1)$.
As discussed already, if $G_Z=0$, a plausible solution would be $G_S=0$
since typically $|G_{SS}|\gg 1$ for the dilaton superpotential induced
by gaugino condensation.
Here it should also be noted that the potential can avoid a run-away
behavior of $S$ with just one gaugino condensation provided that
the uplifting sector leaves a constant superpotential.
This indicates that the uplifting sector can play a role in stabilizing $S$
as pointed out in \cite{Lowen:2008fm}.
Now turning on $G_Z$ which is proportional to $F^Z$, one can find that
the vanishing vacuum energy requires $G_Z={\cal O}(1)$ while $G_S$ is
suppressed compared to $G_Z$ unless the mixing $G_{SZ}$ is unnaturally large.
Hence $F^Z$ is the dominant source of supersymmetry breaking and provides
the uplifting effects which account for the slight shift of the dilaton vacuum
configuration from $G_S=0$.
In fact, the situation is qualitatively same as in usual uplifting
scenarios that have been considered within the framework of type IIB string
theory \cite{Kachru:2003aw,Choi:2004sx,GomezReino:2006dk}.

Let us now examine how the potential for $T$ is affected by the presence of
the F-term uplifting sector and then the possibility to develop a minimum
in $T$ at one of its self-dual points.
For this, it is convenient to change the field basis as
$Z^\prime=\eta(T)^{2n_Z}Z$
so that the uplifting matter field does not transform under the modular
transformations.
The effective superpotential is then generically written as
\bea
\label{WF}
W &=& \frac{1}{\eta(T)^6}\omega(S,Z^\prime),
\eea
as can be inferred from the modular invariance of $G$.
In order for the self-dual point to correspond to a minimum of the potential,
it is simply required to satisfy $\partial_T\partial_{\bar T}V>|\partial^2_T V|$
in the new basis, because
$\partial_T\partial_IV=\partial_T\partial_{\bar I}V=0$ for $I=S,Z^\prime$.
For $V=0$, the mass matrix components evaluated at self-dual points read
\bea
\partial_T\partial_{\bar T} V &\simeq& \frac{3e^G}{(T+T^*)^2}
\left( 1- n_Z + |\lambda(T)|^2 \right),
\nonumber \\
\partial^2_T V &\simeq& \frac{3e^G}{(T+T^*)^2}
(2-n_Z)\lambda(T),
\eea
for $|Z|\ll 1$ and ${\rm Re}(S)={\cal O}(1)$, where
$\lambda(T)=3/2-2(T+T^*)^2\partial^2_T\eta/\eta$, and small corrections
suppressed by $ZG_Z$ and $SG_S$ have been neglected since $G_Z={\cal O}(1)$
and $|G_S|\ll 1$.
Finally, we find that the condition
$\partial_T\partial_{\bar T}V>|\partial^2_T V|$ requires the modular weight
of uplifting field to satisfy
\bea
\label{T-constraint}
T=1 &\quad:\quad& -0.6 \lesssim n_Z \lesssim 2.6,
\nonumber \\
T=e^{i\pi/6} &\quad:\quad& n_Z \lesssim 1,
\eea
in order for the corresponding self-dual point to be a local minimum of the
potential at $V=0$.
It should be noted that the above constraint on $n_Z$ does not depend on the
detailed form of the superpotential of $Z$.
The matter modular weight is known to have a rational number of ${\cal O}(1)$
in orbifold models \cite{Ibanez:1992hc}.
For example, if the uplifting field arises from untwisted sector,
the corresponding modular weight is given by $n_Z=1$ for which the condition
for a minimum can be satisfied at both $T=1$ and $T=e^{i\pi/6}$.

Therefore the uplifting sector plays an important role also in stabilizing
the volume modulus.
Provided that the uplifting field has a modular weight satisfying the
constraint (\ref{T-constraint}), it is indeed possible to get a minimum of the
potential along the $T$-direction at its self-dual points.
Fixed at a self-dual point, the volume modulus has a vanishing F-term and thus
gives contributions neither to the vacuum energy nor soft supersymmetry breaking
terms.
This is phenomenologically desirable as $T$-mediated soft terms are generically
flavor-dependent.
It is also found that the volume modulus has a mass comparable to the gravitino
mass, whereas the dilaton is much heavier than the gravitino since its
non-perturbative superpotential yields a large supersymmetric mass.

\section{Supersymmetry breaking}

In this section, we briefly discuss the pattern of soft supersymmetry
breaking terms arising in the moduli stabilization with F-term uplifting.
The matter part of effective supergravity action is given by
\bea
\label{L-matter}
{\cal L}_{\rm matter} &=&
\int d^4\theta C\bar C Y_i |Q^i|^2
+ \left(\int d^2\theta C^3 \frac{1}{6} \lambda_{ijk} Q^iQ^jQ^k
+ {\rm h.c.} \right),
\eea
where $Q^i$ denote matter superfields, and $C$ is the chiral compensator for
the super-Weyl invariance.
The perturbative shift symmetry associated with ${\rm Im}(S)$ ensures that
holomorphic Yukawa couplings $\lambda_{ijk}$ are independent of $S$, while
the modular invariance constrains their dependence on $T$.
Furthermore, we can simply assume that $\lambda_{ijk}$ are $Z$-independent by
imposing an appropriate symmetry.
On the other hand, the effective K\"ahler potential unavoidably involves
non-renormalizable cross-couplings between $Q^i$ and $Z$, to which sfermion
masses are particularly sensitive, unless they are sequestered from each other.
Taking into account such cross-couplings, the matter wave function is
written as
\bea
\label{Yi}
Y_i &=& (T+\bar T)^{1-n_i}(S+\bar S)^{1/3}\left(1
+ \frac{\epsilon_i}{(T+\bar T)^{n_Z} } |Z|^2 \right),
\eea
for the tree-level K\"ahler potential, where $n_i$ is the modular weight
and the cross-coupling $\epsilon_i$ is a moduli-independent constant.
The structure of sfermion soft terms is essentially determined by how $Y_i$
depends on the supersymmetry breaking fields.

In the case that $T$ is stabilized at a self-dual point, the volume modulus
does not play a role of supersymmetry breaking messenger since $F^T=0$.
Combined with the condition for vanishing vacuum energy, the stationary
condition (\ref{VS0}) then leads to $G_Z={\cal O}(1)$ and
$|G_S|\ll 1$ since the dilation potential is non-perturbatively generated
by gaugino condensation.
Hence the F-terms of $S$ and $Z$ are quite different in size from each other
\bea
\frac{F^Z}{m_{3/2}} = {\cal O}(1), \quad\quad
\frac{1}{m_{3/2}}\left|\frac{F^S}{S}\right| \ll 1,
\eea
in a region with ${\rm Re}(S)={\cal O}(1)$, where $m_{3/2}=e^{G/2}$ denotes
the gravitino mass.
Particularly, this indicates that the dilaton mediation can be comparable
in size to the anomaly mediation \cite{Randall:1998uk}
which is always present and arises at loop level by the F-term of chiral
compensator, $F^C/C\simeq m_{3/2}$.
In such case, one can find that gaugino masses take a mirage pattern since
the gauge coupling functions are determined by the dilaton with
the universal form, $f_a=S$ at the gauge coupling unification scale
$M_{\rm GUT}$\footnote{Due to the cross-couplings in (\ref{Yi}), gaugino
masses receive a non-universal contribution from $F^Z$ at loop level
as a result of the Konishi anomaly \cite{Choi:2007ka}.
This piece is of order $\epsilon_i Z F^Z/8\pi^2$ and thus negligibly small
for $|Z|\ll 1$ compared to the anomaly mediated contribution.}.

The characteristic feature of mirage mediation is the appearance of a mirage
messenger scale $M_{\rm mir}$ \cite{Choi:2005uz}:
\bea
M_{\rm mir} &=& M_{\rm GUT}\left(\frac{m_{3/2}}{M_{Pl}}\right)^{\alpha/2},
\quad\quad
\alpha \,=\, \frac{1}{\ln(M_{Pl}/m_{3/2})}\frac{F^C/C_0}{M_0},
\eea
where $M_0=F^S/(S+S^*)$ denotes the universal dilaton-mediated gaugino mass
at $M_{\rm GUT}$, and $\alpha$ parameterizes the relative size of anomaly
mediation.
For the simple case with a single gaugino condensation, the dilaton
stabilization typically leads to
$F^S/S={\cal O}(m_{3/2}/\ln(M_{Pl}/m_{3/2}))$,
thereby yielding $\alpha={\cal O}(1)$, regardless of the detailed form of
uplifting sector\footnote{In \cite{Lowen:2008fm},
they considered a Polony-like superpotential to describe the supersymmetry
breaking dynamics in the uplifting sector, and found that the dilaton is
stabilized to yield $F^S/S\sim m_{3/2}/\ln(M_{Pl}/m_{3/2})$.}
\cite{Choi:2005uz,Choi:2006xt}.
In addition, one can always make both $\alpha$ and $M_0$ to be real
by using $U(1)_R$ and the perturbative shift symmetry associated with
${\rm Im}(S)$ \cite{Choi:2004sx}.
The mirage mediation scheme shows the mirage unification of gaugino masses
at $M_{\rm mir}$, $M_a(M_{\rm mir})=M_0$, while the gauge couplings are
still unified at $M_{\rm GUT}$.
This reflects that anomaly mediated contributions precisely cancel the
renormalization group evolved parts of gaugino masses at the mirage
messenger scale.

In mirage mediation, low energy A-parameters and sfermion masses are parameterized
in terms of
\bea
\tilde A_{ijk} &=& -F^I\partial_I
\ln\left(
\frac{\lambda_{ijk}}{Y_i(M_{\rm GUT})Y_j(M_{\rm GUT})Y_k(M_{\rm GUT})}
\right),
\nonumber \\
\tilde m^2_i &=& -F^I F^{J*}\partial_I \partial_{\bar J}
\ln Y_i(M_{\rm GUT}),
\eea
where $I$ denote the supersymmetry breaking chiral superfields, i.e. $S$ and $Z$.
As was noticed in \cite{Choi:2005uz}, a mirage pattern is expected for
A-parameters if $\tilde A_{ijk}=M_0$ for non-negligible Yukawa coupling
$\lambda_{ijk}$.
If the condition $\tilde m^2_i + \tilde m^2_j + \tilde m^2_k = M^2_0$ is additionally
satisfied for $\lambda_{ijk}$, sfermion masses also take the mirage pattern.
Provided that above two conditions are satisfied for non-negligible Yukawa
couplings, sfermion soft parameters renormalized at $M_{\rm mir}$ read
\bea
A_{ijk}(M_{\rm mir}) = \tilde A_{ijk}, \quad\quad
m^2_i(M_{\rm mir}) = \tilde m^2_i.
\eea
This indicates that flavor-independent $\tilde m^2_i$ would lead
to the mirage unification of the squark and slepton masses at $M_{\rm mir}$.
For the effective action (\ref{L-matter}), one can easily find
\bea
\tilde A_{ijk} &=& M_0
+ \frac{\epsilon_i + \epsilon_j + \epsilon_k}{(T+T^*)^{n_Z}}Z^*F^Z,
\nonumber \\
\tilde m^2_i &=& \frac{1}{3}M^2_0 - \frac{\epsilon_i}{(T+T^*)^{n_Z}} |F^Z|^2,
\eea
for $\lambda_{ijk}$ having no dependence on $S$ and $Z$.
This shows that A-parameters can take the mirage pattern for
$|Z|\ll 1/8\pi^2$.
On the other hand, sfermion masses are generically dominated by
the contribution from $F^Z={\cal O}(m_{3/2})$ unless
$|\epsilon_i|\ll 1/(8\pi^2)^2$.
In the limit of vanishing cross-couplings, which corresponds to the case
that $Z$ is sequestered from the visible matter superfields, sfermion masses
can share the mirage pattern since
$\tilde m^2_i + \tilde m^2_j + \tilde m^2_k = M^2_0$ .

\section{Effects of anomalous $U(1)_A$ gauge symmetry}

In this section, we examine the effects of anomalous $U(1)_A$ gauge symmetry
on soft supersymmetry breaking terms.
Anomalous $U(1)_A$ gauge symmetry often appears in effective supergravity
derived from heterotic compactifications, where the apparent anomaly is removed
by the Green-Schwarz mechanism \cite{Green:1984sg}.
To implement this mechanism of anomaly cancellation, the dilaton transforms
non-linearly under $U(1)_A$ transformations
\bea
S &\rightarrow& S - \frac{i}{2}\Lambda(x)\delta_{\rm GS},
\eea
where $\Lambda(x)$ is the transformation function and $\delta_{\rm GS}$ is the
Green-Schwarz coefficient of ${\cal O}(1/8\pi^2)$.
The non-linear transformation of $S$ leads to a field-dependent Fayet-Iliopoulos
(FI) term in the effective lagrangian
\bea
\xi_{\rm FI} &=& \frac{1}{2}\delta_{\rm GS} \partial_S K.
\eea
In order to cancel this large FI D-term, it is necessary to introduce a hidden
matter field $X$ that is charged under the $U(1)_A$ gauge group.
Then, along the D-flat direction, the field $X$ acquires a large vacuum value of
${\cal O}(\sqrt{\xi_{\rm FI}})$ for ${\rm Re}(S)={\cal O}(1)$, and subsequently
$U(1)_A$ is spontaneously broken at a very high energy scale.

For the analysis of $U(1)_A$ breaking, we can simply set the visible matter
superfields as $Q^i=0$.
Including the $U(1)_A$ sector, the K\"ahler potential then takes the form
\bea
K &=& -3\ln(T+\bar T) - \ln(S+\bar S-\delta_{\rm GS}V_A)
+ \frac{|Z|^2}{(T+\bar T)^{n_Z}}
+ \frac{\bar X e^{-2V_A}X}{(T+\bar T)^{n_X}},
\eea
where $V_A$ is the $U(1)_A$ vector superfield, $n_X$ is the modular weight of
$X$, and the $U(1)_A$ charge of $X$ is normalized as $q_X=-1$.
From the K\"ahler potential, the D-term reads
\bea
D_A &=& \xi_{\rm FI} + \frac{|X|^2}{(T+T^*)^{n_X}}
= -\frac{\delta_{\rm GS}}{2(S+S^*)} + \frac{|X|^2}{(T+T^*)^{n_X}},
\eea
in which a large value of $\xi_{\rm FI}$ should be cancelled in order to get
a vanishing vacuum energy.
As a result, $U(1)_A$ is spontaneously broken at $|X|$ and the vector gauge
boson acquires a superheavy mass
\bea
M^2_A &=& 2g^2_A \left(\frac{\delta^2_{\rm GS}}{4(S+S^*)^2}
+ \frac{|X|^2}{(T+T^*)^{n_X}} \right),
\eea
where $g_A$ is the $U(1)_A$ gauge coupling.
From the D-flat condition, one can easily find that the $U(1)_A$ gauge boson has
a mass dominated by the contribution from $|X|^2$, and thus its longitudinal
component comes mostly from $X$ rather than from the dilaton.
In fact, the mass eigenstate vector superfield $\tilde V_A$ is given by
\bea
\tilde V_A &\simeq& V_A - \ln|X|,
\eea
while $S$ remains as a flat direction of the D-term potential.
Below the $U(1)_A$ breaking scale, $\tilde V_A$ can be integrated out to
construct an effective supergravity.
The effects of $U(1)_A$ sector are then encoded in effective couplings of the
low energy lagrangian.

Meanwhile, it is possible to estimate the contribution of the $U(1)_A$ sector
to supersymmetry breaking by combining the gauge invariance with the stationary
conditions for $U(1)_A$ charged superfields \cite{Choi:2006bh,Kawamura:1996wn}.
Requiring the gauge invariance of the theory, the stationary conditions
$\partial_{S,X}V=0$ lead to the relations
\bea
\frac{F^X}{X} \simeq -\frac{F^S}{S+S^*},
\qquad
D_A \simeq -\frac{1}{g^2_A}\left|\frac{F^S}{S+S^*}\right|^2,
\eea
for $V=0$, where small corrections suppressed by
$\delta_{\rm GS}={\cal O}(1/8\pi^2)$ have been neglected.
This indicates that $X$-mediation can be as important as the dilaton mediation.
Moreover, if charged under $U(1)_A$, sfermions receive a mass contribution
from the D-term as $\Delta m^2 \sim D_A$, which are also comparable in size
to the dilaton mediated one.
The D-term potential however gives negligible contribution to the vacuum energy
$\Delta V \sim D^2_A$ compared to F-term contributions.
Integrating out the heavy vector superfield, effective couplings of visible
superfields will have a moduli dependence that reflects these features.

To derive an effective theory below the $U(1)_A$ breaking scale, we need to
integrate out heavy $\tilde V_A$ by solving its equation of motion
$\partial K/\partial \tilde V_A=0$ \cite{Choi:2006bh}.
This should be done after including the visible matter superfields.
Requiring $U(1)_A$ gauge invariance, the matter part of supergravity action
is written as
\bea
{\cal L}_{\rm matter} &=&
\int d^4\theta C\bar C Y_i \bar Q^i e^{2q_iV_A} Q^i
+ \left(\int d^2\theta C^3 \frac{1}{6}
\lambda_{ijk} X^{q_i+q_j+q_k}Q^iQ^jQ^k
+ {\rm h.c.} \right),
\eea
where $q_i$ denotes the $U(1)_A$ charge of $Q^i$, the matter wave function
$Y_i$ is given by (\ref{Yi})\footnote{There are in general non-renormalizable
cross-couplings between $X$ and $Q^i$, which are however irrelevant for our
discussion on soft terms since $F^X={\cal O}(X F^S)$ and $|F^S|\ll|F^Z|$.},
and $\lambda_{ijk}$ have no dependence on $S$ and $Z$.
Integrating out the heavy $U(1)_A$ vector superfield using its equation of
motion, the matter part of effective supergravity action is written as
\bea
\label{L-eff-matter}
{\cal L}_{\rm matter} &=&
\int d^4\theta C\bar C Y^{\rm eff}_i |Q^i|^2
+ \left(\int d^2\theta C^3 \frac{1}{6}\lambda^{\rm eff}_{ijk}Q^iQ^jQ^k
+ {\rm h.c.} \right),
\eea
where, under appropriate field redefinition, the effective couplings are given by
\bea
\label{Yieff}
Y^{\rm eff}_i &=& (T+\bar T)^{1-n_i-q_i n_X} (S+\bar S)^{1/3+q_i}
\left(1 + \frac{\epsilon_i}{(T+\bar T)^{n_Z}} |Z|^2 \right)
\left( 1 + {\cal O}(\delta_{\rm GS}) \right),
\nonumber \\
\lambda^{\rm eff}_{ijk} &=& |\delta_{\rm GS}/2|^{(q_i+q_j+q_k)/2} \lambda_{ijk},
\eea
in which $\delta_{\rm GS}={\cal O}(1/8\pi^2)$, and $\epsilon_i$ remains
unchanged since we are considering the case that the uplifting field is not
charged under $U(1)_A$ gauge group\footnote{If the uplifting field is charged
under $U(1)_A$, one can find $D_A={\cal O}(|F^Z|^2/\delta_{\rm GS})$ by
combining the stationary conditions $\partial_{S,X,Z}V=0$ with the gauge
invariance.
In the effective lagrangian, this results in a large cross-coupling between
$Z$ and $Q^i$, $\epsilon_i\sim q_iq_Z/\delta_{\rm GS}$.}.
Note that the moduli dependence of matter wave function is modified in a way
that depends on the $U(1)_A$ charge.
Such change of moduli dependence reflects the effects of $U(1)_A$ sector,
showing that the $U(1)_A$ mediation generates sfermion soft terms which are
comparable in size to the moduli mediated ones.

As discussed in the previous section, the dilaton stabilization naturally
leads to a mirage mediation scheme for $T$ fixed at a self-dual point.
In the presence of the $U(1)_A$ sector, the gaugino masses still show
the mirage unification at $M_{\rm mir}$ since they are not affected
by the sector.
For the sfermion soft terms, the effective matter wave function (\ref{Yieff})
now gives
\bea
\tilde A_{ijk} &=& (1+q_i+q_j+q_k)M_0
+ \frac{\epsilon_i + \epsilon_j + \epsilon_k}{(T+T^*)^{n_Z}}Z^* F^Z,
\nonumber \\
\tilde m^2_i &=& \left(\frac{1}{3} + q_i \right)M^2_0
- \frac{\epsilon_i}{(T+T^*)^{n_Z}} |F^Z|^2,
\eea
where $M_0=F^S/(S+S^*)$.
Since the Yukawa coupling is given by
\bea
\lambda^{\rm eff}_{ijk} &\sim& \frac{\lambda_{ijk}}{(4\pi)^{q_i+q_j+q_k}},
\eea
it is quite plausible to assume $q_i+q_j+q_k=0$ for a large Yukawa coupling
of order unity.
Provided that $q_i+q_j+q_k=0$ for non-negligible Yukawa couplings
$\lambda^{\rm eff}_{ijk}$, the inclusion of $U(1)_A$ sector does not induce
any shift in both $\tilde A_{ijk}$ and $\tilde m^2_i+\tilde m^2_j+\tilde m^2_k$
while providing $\tilde m^2_i$ with an additional piece proportional to $q_i$.
This indicates that the $U(1)_A$ sector can allow more freedom for the values of
sfermion masses while maintaining the mirage pattern of soft parameters.
Further, one can arrange the $U(1)_A$ charges to be flavor-universal in order
to avoid flavor violations.
Such flavor-universal charge assignment would then lead to the mirage
unification of the squark and slepton masses at $M_{\rm mir}$,
if the associated cross-couplings $\epsilon_i$ are sufficiently small.

\section{Conclusion}

In this paper, we have examined the role of F-term uplifting in stabilizing
moduli within the framework of heterotic string theory.
It is noted that the inclusion of an uplifting sector is crucial for fixing
the volume modulus at one of the self-dual points of a modular invariant
potential while achieving a vanishing vacuum energy.
Since the volume modulus does not participate in supersymmetry breaking
at the self-dual points, the F-term uplifting leads to the dilaton stabilization
which can naturally realize the mirage mediation scenario for a moduli potential
generated by gaugino condensation.
Extending to a more general case involving anomalous $U(1)$ gauge symmetry,
we have also examined its implications for supersymmetry breaking.
The supersymmetry breaking in the $U(1)$ sector is essentially related to how the
dilaton is stabilized because the dilaton is responsible for the Green-Schwarz
mechanism of anomaly cancellation.
It turns out that the $U(1)$ sector generically gives a contribution to sfermion
masses comparable to the dilaton mediated one while maintaining the mirage
mediation pattern of soft terms.

\vspace{5mm}
\noindent{\large\bf Acknowledgments}
\vspace{5mm}

The authors would like to thank K. Choi for helpful discussions.
S.S. is supported by the KRF Grants (KRF-2005-210-C0006 and
KRF-2007-341-C00010) funded by the Korean Government.

\end{document}